\documentclass{PoS}
\usepackage{graphicx}
\usepackage{rotating}
\usepackage{verbatim}
\usepackage{amsmath}
\usepackage{amssymb}
\usepackage{gensymb}
\usepackage[numbers]{natbib}
\usepackage{array}
\usepackage{wrapfig,lipsum,booktabs}
\usepackage{mathrsfs}
\usepackage{mathptmx}
\usepackage[nodisplayskipstretch]{setspace}
\newcommand{\ton}{T_{on}}
\newcommand{\toff}{T_{off}}

\newcommand{\non}{N_{on}}
\newcommand{\noff}{N_{off}}
\newcommand{\dt}{\Delta t}

\title{A new time-dependent likelihood technique for detection of gamma-ray bursts with IACT arrays}

\ShortTitle{Time-Dependent Likelihood}

\author{\speaker{Ori M. Weiner} for the VERITAS Collaboration \thanks{veritas.sao.arizona.edu}\\
        Deparment of Physics, Columbia University, New York, NY 10027, USA\\
        E-mail: \email{omw2107@columbia.edu}}

\abstract{In imaging atmospheric Cherenkov telescope (IACT) arrays, the standard method of statistically inferring the existence of a source is based on the maximum likelihood method of Li\&Ma (1983).  We present a new statistical approach, also based on maximum likelihood theory, which takes into account \textit{a priori} knowledge of the source light curve. This approach is especially useful for observations of rapidly decaying gamma-ray bursts (GRBs). We also discuss results established by using this technique to analyze VERITAS GRB observations.}

\FullConference{The 34th International Cosmic Ray Conference,\\
        30 July- 6 August, 2015\\
        The Hague, The Netherlands}

\begin{document}

\section{Introduction}

\subsection{Imaging Atmospheric Cherenkov Telescope Arrays}

Imaging atmospheric Cherenkov telescope (IACT) arrays are the most sensitive instruments for detection of astrophysical gamma-ray emission at $\sim$1-TeV energies \citep{Holder2014}. They rely on the detection of Cherenkov light produced by particles in extensive air showers that were initiated by energetic astrophysical particles entering the atmosphere. The showers are imaged with multiple telescopes and their incoming directions and energies are reconstructed. Most of the astrophysical particles are hadrons, and while many of them are rejected based on the imaged shower's properties, some appear identical to photon-induced showers and are considered to be irreducible \citep{Sobczynska}. The dominant component in the count rate for most sources is this irreducible hadronic background; this is true even after reconstruction of the particle's incoming direction, to an accuracy of about 0.1 degrees. Thus, statistical tools for analysis must be employed and should be designed for maximal sensitivity in a background-dominated regime. 

The standard statistical method used by IACTs was introduced by Li\&Ma in 1983. In their paper the following experiment is assumed \citep{LiMa}: an on-source observation is made for time $T_{on}$ and the instrument is later shifted to observe a nearby background region (off-source) for time $T_{off}$. A test statistic is derived using maximum likelihood theory, given the ratio of observing times $\alpha = \frac{\ton}{\toff}$ and the number of counts during the observations, $\non$ and $\noff$. 

As IACTs evolved the tools to reconstruct the direction of events, the Li\&Ma test statistic was generalized to the "wobble" analysis \citep{Berge}, where source (ON) and background (OFF) observations are made at the same time, but reconstructed to different spatial regions. The ring background model (RBM) is one of the standard tools for such analysis and has been successful in part due to its symmetrical nature that tends to cancel out systematic uncertainties in the background rate \citep{Berge}. 

There are three current-generation IACT arrays, H.E.S.S. \citep{HESS}, MAGIC \citep{MAGIC}, and VERITAS \citep{Holder}. CTA, a next-generation array with a sensitivity improvement of approximately an order of magnitude is expected to be in operation within a few years \citep{CTA}.

\subsection{Gamma Ray Bursts}

Gamma-ray bursts were discovered accidentally in the 1960s by the Vela satellites \citep{VELA}. While much progress has since been made on understanding their origin, the internal mechanism responsible for their prompt radiation is still largely unknown \citep{Piran}. The prompt emission is generally found in the keV-MeV range, but GeV emission has been observed by Fermi-LAT for several bursts \citep{FermiCatalog}. Studies of Fermi-LAT light curves show an approximate power law decay of GRB fluxes \citep{FermiCatalog}. Various models have attempted to explain the emission, some through synchrotron self-Compton scenarios \citep{Sari} which can involve electrons and positrons heated by a forward shock wave. Some of the proposed mechanisms imply substantial emission in the $\sim$1-TeV energy range (for example, \cite{Beloborodov}). Gamma-ray bursts have been a coveted target for IACT observations, and are considered a primary target. None has been detected by any IACT array as of this date.

The Li\&Ma likelihood method has been the standard approach for analysis of GRB observations \citep{Aune}. It uses the fact that the final count tally is Poisson distributed for both the source and background observations regardless of their particular time behaviour during the observation. Time varying sources can thus be inferred using this method, but any prior information on the time behaviour of these sources, as can be provided by other experimental observations or theoretical predictions, cannot be included. We believe the inclusion of \textit{a priori} temporal information can be an important tool in improving the sensitivity for GRB detection.

Generalizations of the Li\&Ma method have been introduced in the past \citep{Klepser}. They generally address the need to detect extended sources, and may need to rely on instrument response functions (IRFs) such as the point spread function or the energy reconstruction responses. To the best of our knowledge, no attempt has been made to explore \textit{a priori} knowledge of the source light curve. We believe this can be achieved without the difficulties associated with the above methods, especially since the time-stamping of events is much more accurate than the variability of any plausible gamma-ray source.

The purpose of this paper is to discuss such a method. It is introduced as a natural generalization of the ring background model, and as such it is also resistant to systematic uncertainties and does not require detailed modelling of instrument response functions. 

Section 2 begins with a derivation of the Li\&Ma test statistic (2.1), and then derives a simplified form of a time-dependent test statistic (2.2). The simplification relies on the assumption that the hadronic background rate is time-independent. Subsection (2.3) shortly describes a modification of the ring background model designed for adjusting to varying background rates. We believe this is the most robust approach for GRB detection.


Section 3 describes results of an analysis of 8 promising VERITAS observations, along with an explanation for how those were selected. The analysis was performed with the method described in Section 2.

\section{Mathematical Derivation}

\subsection{The Li\&Ma likelihood ratio}

We will briefly derive the Li\&Ma test statistic using average background and signal rates as free parameters instead of counts (as used in the original paper). A more detailed discussion can be found in the original paper \citep{LiMa}. Our choice will serve as a smoother transition to the time dependent test statistic.

In the likelihood model, the OFF counts are only due to an unknown background rate whereas the ON counts are explained by an unknown signal rate in addition to the same background rate.

Defining the time-averaged background and signal rates in relation to the expected number of counts: $\overline{b} \toff = <\noff>$, $(\overline{s}+\overline{b})\ton = <\non>$, the likelihood is given by:

\begin{equation}
\begin{split}
\mathscr{L} &= P\left( \non|<\non> \right)P\left( \noff|<\noff> \right) \\ 
&= \frac{{e^{ - (\overline{s}+\overline{b})\ton }  \left( (\overline{s}+\overline{b})\ton  \right) ^{\non} }}{{\non!}} \frac{{e^{ - \overline{b} \toff }  \left( \overline{b} \toff \right) ^{\noff} }}{{\noff!}} \\
\end{split}
\end{equation}

To derive the null hypothesis likelihood, we simply set the signal rate to 0. In this case the average background rate is $\overline{b_0}$ which completely accounts for all counts observed: 

$\overline{b_0} \toff = <\noff>; \quad \overline{b_0}\ton = <\non>$. 

The null hypothesis likelihood is given by:

\begin{equation}
\begin{split}
\mathscr{L}_0 = \frac{{e^{ - \overline{b_0}\ton }  \left( \overline{b_0}\ton  \right) ^{\non} }}{{\non!}} \frac{{e^{ - \overline{b_0} \toff }  \left( \overline{b_0} \toff \right) ^{\noff} }}{{\noff!}} \\
\end{split}
\end{equation}

We find the maximum likelihood values for the rates by maximizing the likelihood of both the null and signal models: $\overline{b_0} = \frac{\non + \noff}{\ton + \toff}; \quad \overline{b} = \frac{\noff}{\toff}; \quad \overline{s} = \frac{\non}{\ton} - \frac{\noff}{\toff}
$

The likelihood ratio then simplifies into:

\begin{equation}
  \dfrac{\mathscr{L}_0}{\mathscr{L}} = \dfrac{\overline{b_0}^{\non + \noff}}{(\overline{b}+\overline{s})^{\non} \overline{b}^{\noff}}
\end{equation}

Wilks' theorem \citep{Wilks} allows us to describe the behaviour of the null likelihood ratio in the regime of high counting statistics. If the null hypothesis is correct, $\sqrt{-2log{\dfrac{\mathscr{L}_0}{\mathscr{L}}}}$ is distributed as a Gaussian variable with a standard deviation of 1 \citep{Wilks}.

\subsection{Time-dependent signal, time-independent background}




To include arrival-time information,  we divide $T_{on}$ into an arbitrarily large number of equal bins N of time $\Delta t$, such that $N\Delta t = T_{on}$. We will require the likelihood model to assign a probability for the number of counts within each bin independently. This will cause the likelihood to approach 0 as $N \rightarrow \infty$ because it will factor in the chance that the arrival times fall within specific bins, the number of which approaches infinity. This behaviour will cancel out in the likelihood ratio test, and the resulting test-statistic will converge nicely. 

In the limit of large N, each time bin will include either a single event, or no events at all. Each of the time bins is independently Poisson distributed with an expectation value approaching 0 as N increases.  



Let $b$ denote the time-independent background rate, and $s(t)$ denote the time-dependent signal rate. The background rate $b$ is treated as an unknown to be optimized by maximum likelihood, and there may be similar unknowns within $s(t)$, such as the amplitude or, a "shape parameter", etc. For use of Wilks's theorem we must require nested models, and thus $s(t)$ must have at least one such unknown, most simply the amplitude. If use of Wilks's theorem is not possible, computer modelling can replace it, and the condition above can be relaxed.

We denote the arrival times of signal (ON) photons as $\{t_{on}\} = (t_1, t_2 ... t_{n_{on}})$. The likelihood is a product of the Poisson probabilities for the count tally over all N time bins:


\begin{equation}
\begin{split}
\mathscr{L} = &\left(\prod_{t_i = (\Delta t, 2 \Delta t, ... N \Delta t)} \frac{[\Delta t(b+s(t_i))]^{\{0,1\}}}{\{0,1\}!} e^{-\Delta t(b+s(t_i))} \right) \frac{(b\toff)^{\noff}}{\noff !} e^{-b\toff} \\
\end{split}
\end{equation}


where \{0,1\} are chosen depending on whether there is an event in the $t_i$ bin.

\begin{equation}
 \lim_{N \to \infty} \mathscr{L} = \dt^{\non} \left(\prod_{t_i \in \{t_{on}\}} (b+s(t_i)) \right) \frac{(b\toff)^{\noff}}{\noff !} e^{-b(\ton + \toff) - \int_0^{\ton}dt \, s(t)}
\end{equation}

For the null hypothesis, we set $s(t)=0$, and denote the background rate as $b_0$, which will obey essentially the same likelihood ratio as the Li\&Ma null hypothesis, with only a change of constants.

\begin{equation}
\begin{split}
\mathscr{L}_0 = \dt^{\non} b_0^{\non} \frac{(b_0\toff)^{\noff}}{\noff !} e^{-b_0(\ton + \toff)}
\end{split}
\end{equation}

Thus $\mathscr{L}_0$ is also maximized by: $b_0 = \frac{\non + \noff}{\ton + \toff}$, giving:

\begin{equation}
\begin{split}
\mathscr{L}_0 = \dt^{\non} b_0^{\non} \frac{(b_0\toff)^{\noff}}{\noff !} e^{-(\non + \noff)}
\end{split}
\end{equation}

The last equality follows from plugging in the exact form of $b_0$ into the exponential. The likelihood ratio is given by:

\begin{equation}
\begin{split}
 \dfrac{\mathscr{L}_0}{\mathscr{L}} = {} &\dfrac{b_0^{\non + \noff}}{\left(\prod_{t_i \in \{t_{on}\}} (b+s(t_i)) \right) b^{\noff}} e^{b(\ton + \toff) + \int_0^{\ton}dt \, s(t) - (\non + \noff)}
\end{split}
\end{equation}

This ratio can be further simplified by exploring the connection between $s(t)$ and $b$. To do so, we must find the maximum of $\mathscr{L}$ or equivalently of $\log\mathscr{L}$. For the purpose of GRB detection we will leave only one free parameter, the amplitude, in the signal time profile, $s(t) \stackrel{\Delta}{=} \theta f(t) $. This choice reflects the certainty of the flux decaying rapidly (usually as a power law), and the uncertainty about the amplitude of VHE emission.

\begin{equation} \label{eq1}
\begin{split}
\frac{\partial{\log\mathscr{L}}}{\partial{b}} = \frac{\noff}{b} + \sum_{t_i \in \{t_{on}\}}\frac{1}{b+\theta f(t_i)}-(\ton + \toff) = 0 \\
\end{split}
\end{equation}

\begin{equation} \label{eq2}
\begin{split}
\frac{\partial{\log\mathscr{L}}}{\partial{\theta}} &= \sum_{t_i \in \{t_{on}\}}\frac{f(t_i)}{b+\theta f(t_i)}-\int_0^{\ton}dtf(t) = 0 \\
\end{split}
\end{equation}

An important identity is derived by noting that at the maximum of the likelihood function we can assert: $b\frac{\partial{\log\mathscr{L}}}{\partial{b}} + \theta \frac{\partial{\log\mathscr{L}}}{\partial{\theta}}  = 0$.


\begin{equation} \label{eq3}
\begin{split}
b(\ton + \toff) + \int_0^{\ton}dt \, s(t) = \non + \noff
\end{split}
\end{equation}

The above relation leads to the cancellation of the exponential in the test statistic. It also allows a substitution of $b = \frac{\non + \noff - \theta \int_0^{\ton}dtf(t)}{\ton + \toff}$ into equation (2.9), which results in a polynomial equation of order $\non+1$ for $\theta$. It can be solved using a computer grid search, or any other suitable optimization algorithm to find a maximum likelihood $\theta$. Due to the condition above, any grid search will only need to search over values of the amplitude, $\theta$.

We can now write a simplified form for the test statistic:

\begin{equation}
\dfrac{\mathscr{L}_0}{\mathscr{L}} = \dfrac{b_0^{\non + \noff}}{\left(\prod_{t_i \in \{t_{on}\}} (b+\theta f(t_i)) \right) b^{\noff}}
\end{equation}

In the case of a time-independent signal rate, this ratio is equivalent to the Li\&Ma likelihood ratio, when written in terms of average rates. For variable light curves, ON counts with arrival times that match the expected profile of $f(t)$ receive an essentially higher "weight", according to how bright the source was expected to be at the time. 


\subsection{\textit{A priori} time-dependent background} 

The background rate can also be modelled by an \textit{a priori} time behaviour, for example by using a known dependence on the observation zenith angle, or using an independent part of the field of view for this purpose. This generalization of the previous section, based on the Ring Background Model \citep{Berge}, has been prepared and will be fully described in a future paper \citep{Weiner}. It is similar in nature to the technique described in Section 2.2 and is capable of handling a time-varying background rate.



\section{Analysis of VERITAS observations}

In the VERITAS observing program, gamma-ray bursts are considered the highest-priority sources. GCN alerts with a finer than $10 \degree$ localization uncertainty radius are treated as a priority and prompt the observers to slew immediately to any such burst that is above a $20 \degree$ elevation. 

VERITAS sensitivity is expected to be sufficient to detect strong bursts under reasonable assumptions about their high energy spectrum. For example, an extrapolation of LAT observations for GRB 130427A predicts an initial VERITAS flux of about 100 photons per second during the prompt phase \citep{Aune}, while, for comparison, the background rate for a region the size of the point-spread-function (PSF) is on the order of 1 count per minute. At the night of the burst, VERITAS was not operating due to a full moon; however, according to the estimates, a detection of GRB 130427A would have been achieved in less than a second. Observations started the following night, and results yielded a small positive significance of 1.3 for the first night, and 1.1 for the second night \citep{Aune}. 

As of the time this document has been written, VERITAS has observed 132 gamma-ray burst locations. Of those observations, we attempted to analyze the most promising candidates for detection based on a reasonable \textit{a priori} measure. The most important factors in our ability to detect a burst are the observing delay, redshift (z), and elevation angle at the time of observing. The elevation angle is particularly important for high z bursts, since the instrument's energy threshold rapidly increases with decreasing elevation angle, whereas absorption of gamma-rays due to the extragalactic background light (EBL) favours a very soft spectrum. We also required the burst to have been localized as well as our angular resolution or better ($\sim0.1 \degree$). Bursts with unknown redshift were left out of the analysis, considering the high z typically associated with them. 

We calculated the following weight for each burst that met the conditions above: $W = e^{-\tau(z,\theta)}/t_{obs}$, where $\tau$ is the optical depth due to EBL absorption at the threshold energy, which depends on the elevation angle of observation $\theta$, and $t_{obs}$ is the observing delay. One can recognize this expression to be approximately proportional to the incoming VHE flux at the initial time of observing, assuming the flux decays as $1/t$, as is more or less typical for Fermi-LAT observed bursts \citep{FermiCatalog} . The most promising observation was found to be that of GRB 150323A. We analyzed other bursts with weights within a factor of about 100 of its weight, to account for possible intrinsic variance in the VHE emission of the bursts. We found 7 other observations that met our criteria (see Table 1). 

The bursts were analysed using a standard VERITAS analysis package. Standard cuts optimized for soft spectrum sources were used in the analysis. The VHE flux was assumed to decay as a powerlaw with an index of $-1$, beginning at the prompt phase of emission. The results for each of the 8 bursts is described in the Table 1. 

A form of a stacking analysis of the bursts was performed with what we will call the binomial test. It is performed as follows: Each burst is analyzed and its significance is found. A  threshold significance is decided prior to analysis, and each burst can pass a test by having a greater significance than this decided threshold, or otherwise fail the test. A p-value is calculated given the number of bursts that passed the test compared to what would be expected if their significance values followed a unit Gaussian distribution, as one would expect from the null hypothesis.

\begin{wraptable}{r}{5.5cm}
\begin{center}
    \begin{tabular}{ | >{\centering\arraybackslash}p{3cm} | >{\centering\arraybackslash}p{2cm} | }
    \hline
    Name of Burst & Significance \\ \hline
    GRB111225A & -0.693 \\ \hline
    GRB120422A & 2.473 \\ \hline 
    GRB130215A & -0.600 \\ \hline   
    GRB130427A & 1.690 \\ \hline
    GRB130604A &  -0.681 \\ \hline   
    GRB140622A & -1.327 \\ \hline   
    GRB150120A & -0.490 \\ \hline
    GRB150323A & -1.541 \\ \hline 
    \end{tabular}
\end{center}
\caption{Time-dependent likelihood significance of 8 selected VERITAS bursts}
\label{VERITAS - ICRC 2015}
\end{wraptable} 

The p-value is calculated with the following formula and then converted to Gaussian significance:

\begin{equation}
P = \sum_{n=l}^{N} \left( \Phi^n(s) (1-\Phi(s))^{N-n} {n \choose N} \right)
\end{equation}

where $l$ is the number of bursts passing the test, $N$ is the total number of bursts in the analysis, and $\Phi(s)$ is defined here as the one-sided probability of obtaining a result with Gaussian significance greater than $s$: $\Phi(s) = \int_s^{\infty} dx \frac{1}{\sqrt{2\pi}} e^{-\frac{x^2}{2}}$.

The binomial test p-value may also be multiplied by the number of trials. In this analysis, we accounted for 10 trials while using only 3 of them (see Table 2). The reason for this is that we are planning more trials on a wider selection of bursts without a known redshift.

The threshold significance values were carefully selected using a computer script, to make sure that a whole number of bursts will be required to give an exact 5-sigma result. This resulted in fractional values for the thresholds, to a precision of 3 decimal digits. Table 2 summarizes our test selection (thresholds are calculated using a trials factor of 10). Table 2 describes the final p-value of all 3 binomial trials.

\begin{table}[h!]
\begin{center}
    \begin{tabular}{ | >{\centering\arraybackslash}p{2cm} | >{\centering\arraybackslash}p{3cm} | >{\centering\arraybackslash}p{3cm} | >{\centering\arraybackslash}p{2cm} | }
    \hline
    Threshold significance & Minimal number of bursts needed to achieve detection & Number of bursts passing test & p-value \\ \hline
    1.852 & 6 out of 8 & 1 out of 8 & 0.23 \\ \hline
    2.611 & 4 out of 8 & 0 out of 8 & 1 \\ \hline
    3.998 & 2 out of 8 & 0 out of 8 & 1 \\ \hline
    \end{tabular}
\end{center}
\caption{Description and results of all binomial tests}
\label{table:2}
\end{table}


\subsection*{Summary}
We discussed a new technique for detecting time-dependent sources with IACTs; one that we believe can be especially useful for detection of gamma-ray bursts. We used this technique to perform a stacking analysis of 8 promising observations made by VERITAS. No significant detection is found.

\subsection*{Acknowledgements}

This research is supported by grants from the U.S. Department of Energy Office of Science, the U.S. National Science Foundation and the Smithsonian Institution, and by NSERC in Canada. We acknowledge the excellent work of the technical support staff at the Fred Lawrence Whipple Observatory and at the collaborating institutions in the construction and operation of the instrument. The VERITAS Collaboration is grateful to Trevor Weekes for his seminal contributions and leadership in the field of VHE gamma-ray astrophysics, which made this study possible. We thank Lydia Seymour for assisting us in compiling information on VERITAS GRB observations.

\end{document}